\documentclass{aa}
\usepackage[varg]{txfonts}
\usepackage{xcolor}
\usepackage{hyperref}
\usepackage{natbib}
\hypersetup{
    colorlinks,
    linkcolor={red!80!black},
    citecolor={blue!80!black},
    urlcolor={blue!80!black}
}
\bibpunct{(}{)}{;}{a}{}{,}

\begin{document}

\title{Directly comparing coronal and solar wind elemental fractionation}
\author{
	D. Stansby \inst{1}
	\and D. Baker \inst{1}
	\and D. H. Brooks \inst{2}
	\and C. J. Owen \inst{1}
}
\institute{
	Mullard Space Science Laboratory, University College London, Holmbury St. Mary, Surrey RH5 6NT, UK \\ \email{d.stansby@ucl.ac.uk} \label{inst1}
	\and College of Science, George Mason University, 4400 University Drive, Fairfax, VA 22030, USA \label{inst2}
}

\abstract{As the solar wind propagates through the heliosphere, dynamical processes irreversibly erase the signatures of the near--Sun heating and acceleration processes. The elemental fractionation of the solar wind should not change during transit, however, making it an ideal tracer of these processes.
}
{We aim to verify directly if the solar wind elemental fractionation is reflective of the coronal source region fractionation, both within and across different solar wind source regions.
}
{
A backmapping scheme was used to predict where solar wind measured by the \emph{Advanced Composition Explorer} (ACE) originated in the corona. The coronal composition measured by the \emph{Hinode} \emph{Extreme ultraviolet Imaging Spectrometer} (EIS) at the source regions was then compared with the in-situ solar wind composition.
}
{
On hourly timescales, there is no apparent correlation between coronal and solar wind composition. In contrast, the distribution of fractionation values within individual source regions is similar in both the corona and solar wind, but distributions between different sources have a significant overlap.
}
{
The matching distributions directly verify that elemental composition is conserved as the plasma travels from the corona to the solar wind, further validating it as a tracer of heating and acceleration processes. The overlap of fractionation values between sources means it is not possible to identify solar wind source regions solely by comparing solar wind and coronal composition measurements, but a comparison can be used to verify consistency with predicted spacecraft-corona connections.
}
\keywords{Sun: abundances -- Sun: corona -- Sun: heliosphere -- solar wind}
\maketitle
\section{Introduction}

The solar wind is a stream of plasma, originating at the surface of the Sun and flowing out via the corona to fill interplanetary space. Despite a long history of remotely observing the Sun and locally sampling the solar wind, the source regions, acceleration, and heating mechanisms of the solar wind are still not comprehensively known \citep[e.g.~see][for recent reviews]{Abbo2016, Cranmer2017}.

In terms of number density, the solar wind is primarily composed of fully ionised hydrogen and helium, but a host of trace positive heavy ions are also present \citep{Bame1975, Bochsler2007}. Since these ions form a super-sonic beam of particles flowing away from the Sun \citep{Ogilvie1980, Bochsler2007} and as their collisional frequency is very small \citep{Hundhausen1968}, the relative elemental and charge fractionation of the solar wind should be preserved during transit. Plasma processes occurring close to the Sun, which cause variations in the heavy ion fractionation, can therefore be inferred from these in-situ measurements, even when taken far away from the Sun.

The relative abundance of different elements is the same everywhere in the photosphere \citep{Asplund2009}, meaning any variances in solar wind abundances must be due to processes happening above the photosphere, either in the chromosphere or corona. Indeed, remote sensing measurements of the corona reveal a spatial variation of elemental abundances: in active regions, elements that are more easily ionised (low first ionisation potential (FIP) elements) are more abundant than those that are less easily ionised (high FIP elements) \citep[e.g.][]{Meyer1985, Uzzo2004, Brooks2011, Baker2018, Doschek2019}. A similar overabundance of high FIP elements is also seen in coronal streamers \citep[e.g.][]{Ko2002, Uzzo2006, Uzzo2007}. In contrast, no such enhancements are seen above coronal holes \citep{Feldman1998}.

The enhancement of low FIP elements can be explained by the existence of an upward force that is exerted only on ionised elements.  Low FIP elements are more easily ionised; therefore in areas where an upward force is acting, their vertical flux relative to high FIP elements is enhanced, causing a relative over-abundance on the low FIP elements \citep[e.g.][]{Vauclair1996, Arge1998, Diver2005, Laming2004, Laming2017}. This fractionation takes time to build up, typically 2 to 3 days \citep{Widing2001, Baker2018}, so plasma must be confined in the chromosphere or corona over these timescales for this fractionation to develop \citep[see][for a review of the FIP effect]{Laming2015}. As a result, plasma on closed active region loops tends to fractionate, but plasma on open coronal hole field lines that rapidly escapes the corona does not \citep{Geiss1995, Steiger2000}.

Previous attempts to track coronal compositional signatures into the solar wind have primarily focused on active regions, using magnetic field modelling to assess whether outflows measured in the corona were on open field lines that extended into the solar wind. Only some active regions contain open magnetic field lines \citep{Edwards2016}, but of those that do, the remote compositional signatures sometimes match the solar wind compositional signature measured a few days later at 1~AU \citep{Brooks2011, vanDriel-Gesztelyi2012}. Recently, more detailed mapping has confirmed the match in compositional signatures from a handful of isolated active region solar wind sources \citep{Slemzin2013, Culhane2014, Macneil2019a}. Compositional signatures have also been successfully tracked from above the streamer belt into the solar wind \citep{Bemporad2003}. Although the compositional signatures of isolated sources have been directly matched between the corona and solar wind, as far as we are aware there has not yet been a study of whether changes in composition are traceable, either within a single source region or across multiple source regions.

In this paper we preform these studies using a full disc observation of the Sun taken by the \emph{Extreme ultraviolet Imaging Spectrometer} (EIS) on board \emph{Hinode}, and a subsequently derived full Sun composition map. This map was first used by \cite{Brooks2015} to identify the source regions of slow solar wind, and estimate their collective mass flux from the coronal measurements. We expanded on this study by using magnetic field modelling to estimate source regions of the solar wind measured by the ACE spacecraft at 1 AU. We found 3 distinct source regions measured by the \emph{Advanced Composition Explorer} (ACE), and present in the full Sun composition map. We then compared the composition variations both within and between these sources in both the solar wind and the corona.

Section \ref{sec:data} describes the data used in this study, and Sect.~\ref{sec:mapping} presents the backmapping procedure used. In Sect.~\ref{sec:comparison} the direct comparison between coronal and solar wind composition is presented, and Sect.~\ref{sec:conclusions} concludes and discusses our results, which have a particular relevance for the recently launched \emph{Solar Orbiter} mission.

\section{Data}
\label{sec:data}
\subsection{Remote sensing}
\subsubsection{Coronal composition}
In order to measure elemental fractionation in the corona, a full disc set of observations taken by the EIS instrument \citep{Culhane2007} aboard the \emph{Hinode} spacecraft \citep{Kosugi2007} were used\footnote{Data available at \url{http://solar.ads.rl.ac.uk/MSSL-data/eis/level2/}}. The scans were carried out from the 16 -- 18 Jan 2013, and were first analysed by \cite{Brooks2015}. Over the three days, 26 individual rasters were taken to build up a full Sun view. The wavelengths measured were primarily iron emission lines\footnote{Fe VIII 185.213 \AA, Fe IX 188.497 \AA, Fe X 184.536 \AA, Fe~XI~188.216~\AA, Fe XI 188.299 \AA, Fe~XII~195.119~\AA, Fe~XII~203.72~\AA, Fe XIII 202.044 \AA, Fe XIII 203.826 \AA, Fe XIV 264.787 \AA, Fe~XV~284.16~\AA, and Fe XVI 262.984 \AA} that can be used to determine electron density and the differential emission measure (DEM), along with individual silicon and sulphur emission lines\footnote{Si X 258.37 \AA~and S X 264.22 \AA}, which can be used to make elemental abundance measurements.

The electron density and DEM estimations were used to model the intensity ratio of the Si and S lines, giving the silicon to sulphur abundance ratio ($n_{Si} / n_{S}$). For a full account of the method, see \cite{Brooks2015}. When normalised to the Si/S photospheric abundance ratio of 2.34 \citep{Scott2015a}, this ratio is used as a proxy for the First Ionisation Potential (FIP) bias ratio. The map of FIP bias ratios is shown later in Fig.~\ref{fig:fip_map}. Although sulphur is commonly used as a high FIP element in studies, it lies close to the boundary between low and high FIP elements, and it has been suggested that it can exhibit both low and high FIP behaviour \citep{Reames2018, Kuroda2020}. Ideally another element with a higher FIP would be used, but emission lines from other high FIP elements are challenging to measure with \emph{Hinode}/EIS \citep{Feldman2009}.

\subsubsection{Context}
\label{sec:context}
\begin{figure*}
	\includegraphics[width=2\columnwidth]{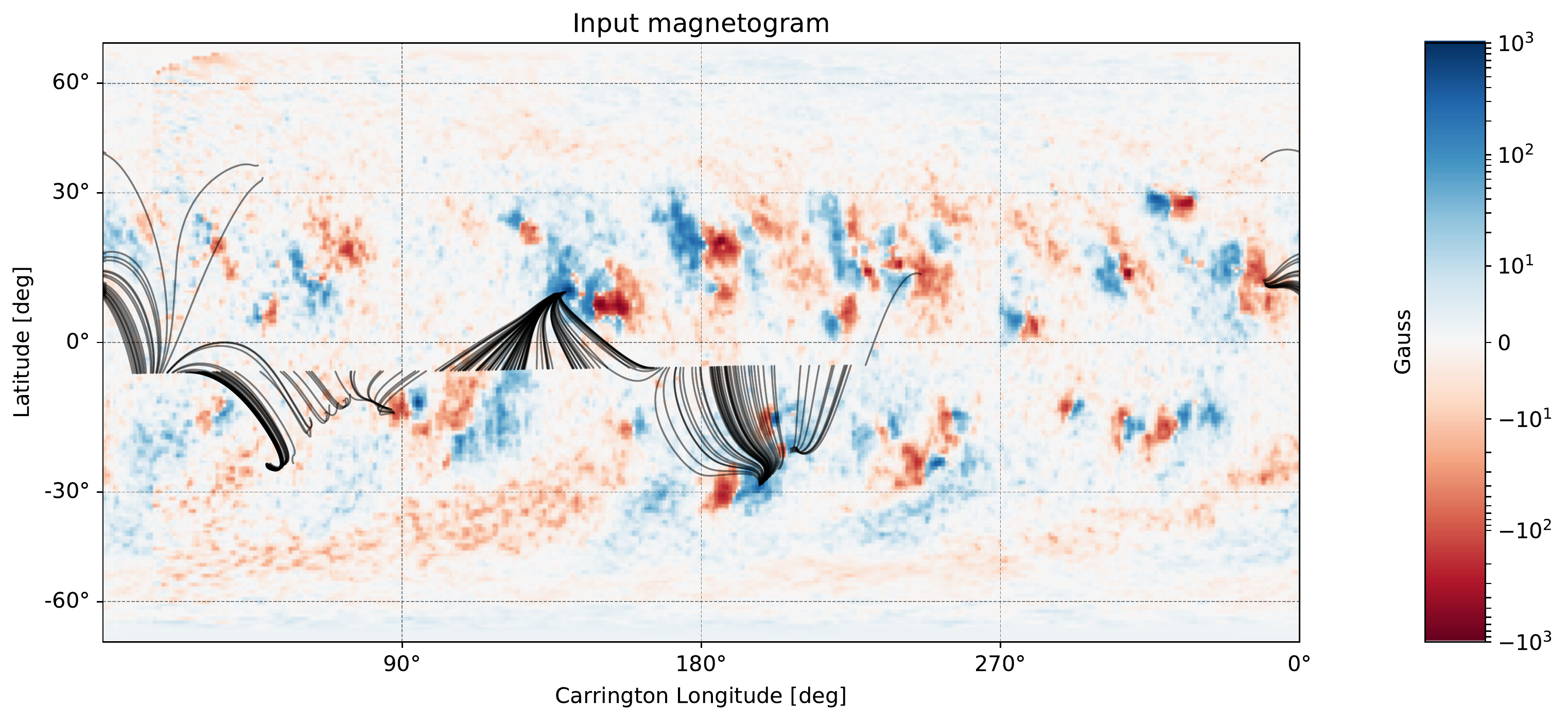}
	\includegraphics[width=2\columnwidth]{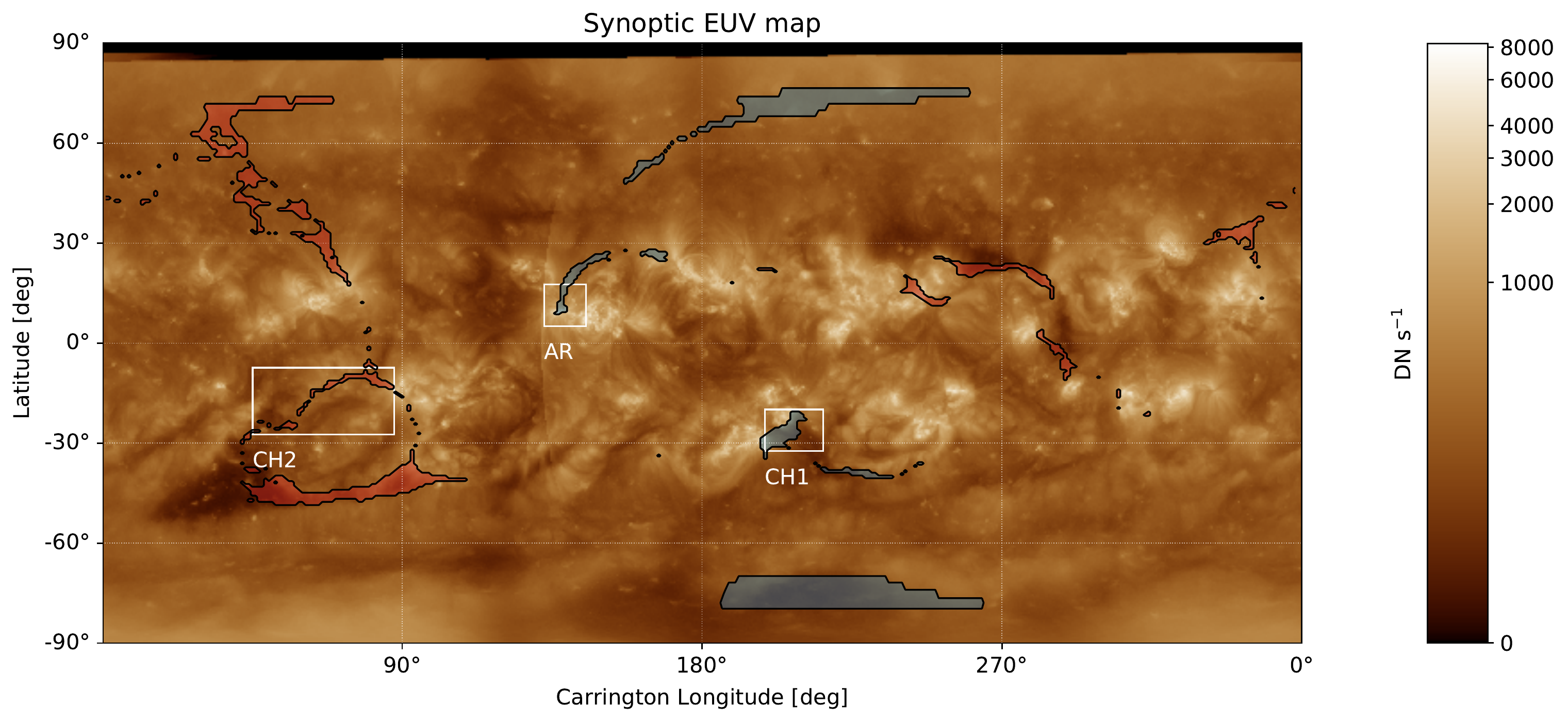}
	\caption{Overview of magnetic field mapping. Top panel shows the GONG radial field magnetogram used as input to PFSS modelling. Black lines show magnetic field lines traced through the PFSS solution from the source surface to the solar surface. Bottom panel shows a synoptic AIA 195\AA~EUV map, with overplotted contours showing the open field regions of the PFSS model. The white and associated labels identify the three solar wind source regions discussed in the text.}
	\label{fig:ss_mapping}
\end{figure*}

As part of the backmapping scheme a global Potential Field Source Surface (PFSS) model of the coronal magnetic field at the time of the EIS observations was used. As the input to this model, a solar surface line-of-sight magnetogram measured by the \emph{Global Oscillations Network Group} (GONG) consortium was used, which was last updated on 20 Jan 2013\footnote{Available at \url{ftp://gong2.nso.edu/oQR/zqs}}. Although the solar magnetic field evolved over the $\sim$half a solar rotation of in-situ data studied here, the open and closed field regions predicted by different maps measured over this time span did not vary significantly, so for simplicity a single synoptic map was used. This is shown in Fig.~\ref{fig:ss_mapping}, and discussed later in Sect.~\ref{sec:mapping}.

For visual context, extreme ultra-violet (EUV) images at 193\r{A} were taken\footnote{Available at \url{http://jsoc.stanford.edu/ajax/exportdata.html}} from the \emph{Atmospheric Imaging Assembly} \citep[AIA,][]{Lemen2012} on board the \emph{Solar Dynamics Observatory} \citep[SDO,][]{Pesnell2012}. These EUV images show emission primarily from Fe XII and XXIV in the corona, and reveal coronal holes (areas of dark emission with open field lines), quiet Sun areas (areas of intermediate emission with weak magnetic field footpoints and closed field lines), and active regions (areas of bright emission, with strong magnetic field footpoints that may be open or closed). A full sun EUV map for the interval of this study is shown in Fig.~\ref{fig:ss_mapping}.
\subsection{In situ}
The solar wind in-situ data used were measured by the ACE spacecraft \citep{Stone1998}, which was located at the Sun-Earth L1 point, approximately 1 AU away from the Sun. Heavy ion composition measurements were taken from the \texttt{swi\_h3b} dataset\footnote{Available at \url{https://cdaweb.sci.gsfc.nasa.gov/misc/NotesA.html\#AC_H3_SW2}} measured by the \emph{Solar Wind Ion Composition Spectrometer} \citep[SWICS,][]{Gloeckler1998}.

After August 2011 the SWICS instrument underwent re-configuration, which meant the in-situ silicon to sulphur ratio (Si/S) was not available for direct comparison with the coronal measurements. Instead the in-situ iron to oxygen ratio (Fe/O) was used, which is expected to show similar trends to Si/S, even if the absolute magnitudes of the ratios differ, as they both represent the ratio of a low to high FIP element.

During data analysis we found that the provided SWICS Fe/O values were systematically higher after the re-configuration compared to before. We therefore manually corrected the more recent values to match the distribution of pre-re-configuration values, when SWICS was operating as originally intended. Details of the correction are given in Appendix 1. Similarly to the Si/S ratio, the Fe/O ratio can be normalised by its photospheric abundance ratio of 0.13 \citep{Scott2015} to give another proxy for the FIP bias.
In addition to heavy ion data, in-situ magnetic field measurements from the ACE \emph{Magnetic Fields Experiment} \citep[MFI,][]{Smith1998} were used to measure the local magnetic polarity, and proton solar wind speed measurements from the ACE \emph{Solar Wind Electron Proton Alpha Monitor} \citep[SWEPAM,][]{McComas1998} were used for ballistic backmapping.
\section{Results}
\subsection{Identifying solar wind source regions}
\label{sec:mapping}

In order to assign source regions for the solar wind measured in-situ by ACE at 1 AU, a common two step backmapping procedure was used \citep[e.g.][]{Neugebauer1998}. The solar wind was mapped ballistically back to a source surface at 2.3R$_{\odot}$ using the radial proton velocity measured in-situ by ACE. Although a simple ballistic backmapping scheme can cause un-physical streamline crossover \citep[e.g.][]{Riley2011}, checks showed this was minimal and the source surface Carrington longitude almost always increased as a function of time.

Between 2.3R$_{\odot}$ and 1R$_{\odot}$ a potential field source surface \citep[PFSS,][]{Schatten1969, Altschuler1969} model was computed, using the \emph{pfsspy} software package \citep{Yeates2018, Stansby2020c}. The GONG magnetogram was used as the lower boundary condition, with the solution calculated at the 360$^{\circ}$ $\times$ 180$^{\circ}$ resolution of the input magnetogram on 60 radial grid points. Below the source surface, the solar wind was assumed to flow directly along magnetic field lines traced through the PFSS solution to the solar surface. The source surface height of 2.3R$_{\odot}$ was chosen empirically to maximise the coverage of coronal holes in the EUV image by open field regions as predicted by the PFSS model.

The results of this mapping over an 18-day period are shown in Fig.~\ref{fig:ss_mapping}, with the input magnetogram and traced PFSS field lines (top panel), and an open-closed field map overlaid on an AIA 193\AA~synoptic map (bottom panel).  The solar footpoint of the traced field lines as a function of time transitions from right to left in the Carrington frame of reference used for the synoptic maps. The footpoint at the start of the interval rested in a positive polarity coronal hole region south of equator at around 200$^{\circ}$ longitude, labelled as CH1 near the middle of Fig.~\ref{fig:ss_mapping}. From there it crossed north of the equator to a positive polarity active region at around 130$^{\circ}$ longitude (labelled as AR in Fig.~\ref{fig:ss_mapping}), and then crossed the polarity inversion line and connected to a small negative polarity active region at 90$^{\circ}$ longitude before transitioning to an adjacent thin coronal hole from 90$^{\circ}$ - 45$^{\circ}$ longitude, labelled as CH2 on the left-hand side of Figure \ref{fig:ss_mapping}.

\begin{figure}
	\includegraphics[width=\columnwidth]{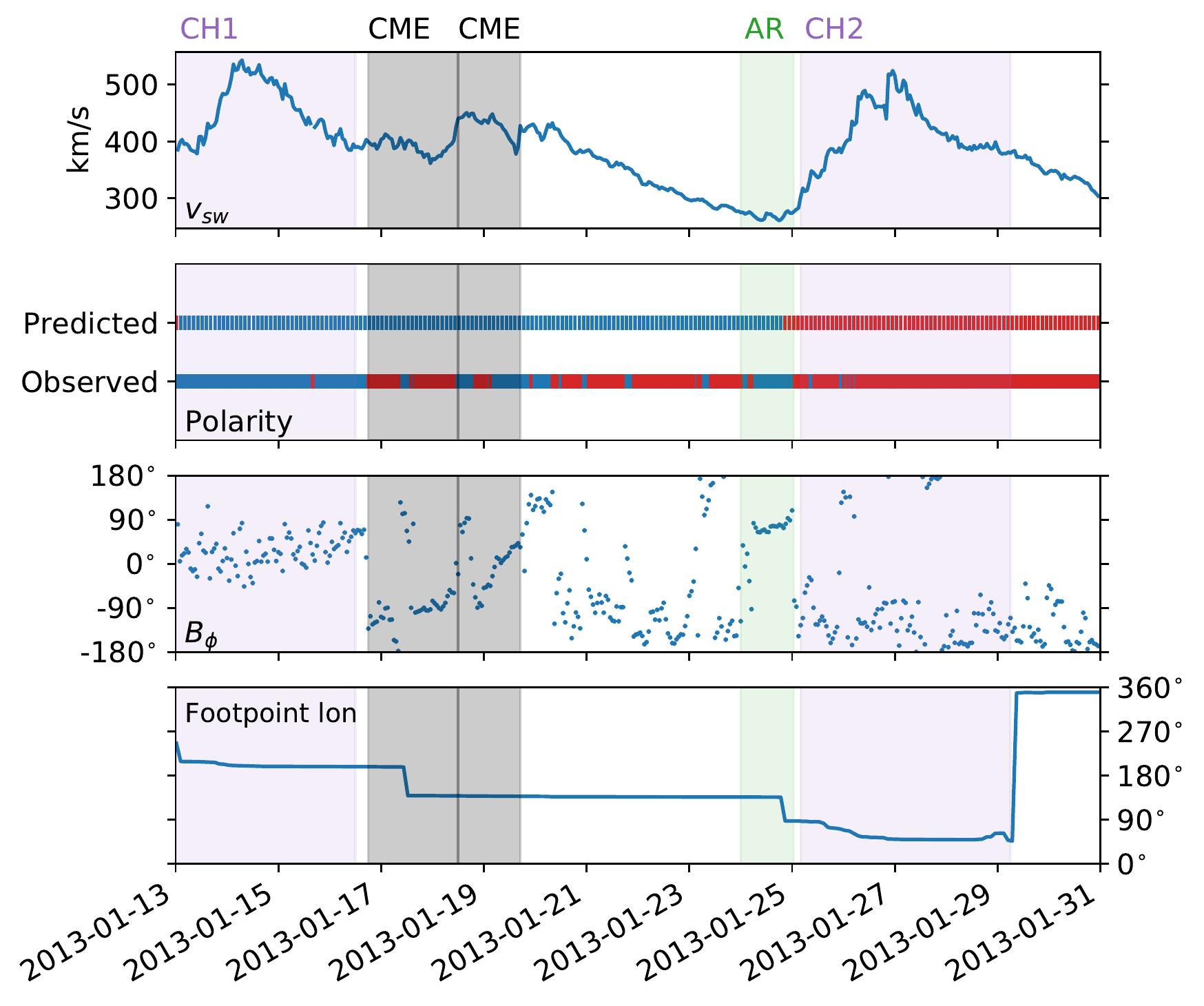}
	\caption{Overview of ACE solar wind measurements from 4 Jan 2013 to 31 Jan 2013. From top to bottom, solar wind speed, predicted (from the PFSS model) and observed (from in-situ measurements) magnetic field polarity, in-situ magnetic field azimuth, and mapped solar footpoint Carrington longitude.}
	\label{fig:L1polarity}
\end{figure}
To verify the mapping from the solar surface to ACE, Fig.~\ref{fig:L1polarity} shows a comparison between in-situ solar wind properties and the predicted magnetic polarity over the 18-day period. The top panel shows the solar wind speed, with slow wind in the middle of the interval bookended by two faster streams. The predicted polarity (given by sign$(B_{r})$ on the PFSS source surface), and in the in-situ polarity (given by sign$(\mathbf{B} \cdot \mathbf{B}_{spiral})$, where $\mathbf{B}_{spiral}$ is the predicted Parker spiral field \citep{Parker1958}) are shown in the second panel. The in-situ magnetic field clock angle in the ecliptic plane shown in the third panel.

Before 17 Jan 2013 and after 25 Jan 2013 the measured and model polarity agreed, but between these two dates the positive polarity predicted by the PFSS model did not agree with the in-situ measurements. This is most likely because of two CMEs that erupted and arrived at ACE during this time period, listed on the Richardson \& Cane ICME list\footnote{\url{http://www.srl.caltech.edu/ACE/ASC/DATA/level3/icmetable2.htm}}. These are observed as large scale magnetic field rotations in $B_{\phi}$, (Fig.~\ref{fig:L1polarity}, third panel), and indicated by black shaded bars. As well as locally disrupting solar wind structure, the wake of the CMEs presumably also disrupted the trailing solar wind, resulting in the mis-match between predicted and in-situ polarity from 17 Jan 2013 to 23 Jan 2013, and therefore an incorrect connection prediction. For this reason, this period was excluded from further analysis.

The beginning and end of the whole time interval, which both contain relatively faster solar wind, have magnetic polarities that agree. The bottom panel of Fig.~\ref{fig:L1polarity} shows the footpoint longitude as a function of time, with the backmapping predicting footpoints inside coronal holes. This agrees with the measurement of faster wind, giving confidence in the backmappiing for these intervals. These two coronal hole intervals are labelled `CH1' and `CH2', and shown with light purple bars in Fig.~\ref{fig:L1polarity}.

On 24 Jan 2013 there is also an interval of solar wind where the PFSS polarity matches the polarity observed in-situ. We therefore take this as evidence that mapping for this smaller time period is also correct. The mapping for this period points back to an active region outflow, so is labelled `AR', and shown with a light green bar in Fig.~\ref{fig:L1polarity}.

\subsection{Comparing coronal and in-situ properties}
\label{sec:comparison}
\begin{figure*}
	\includegraphics[width=2\columnwidth]{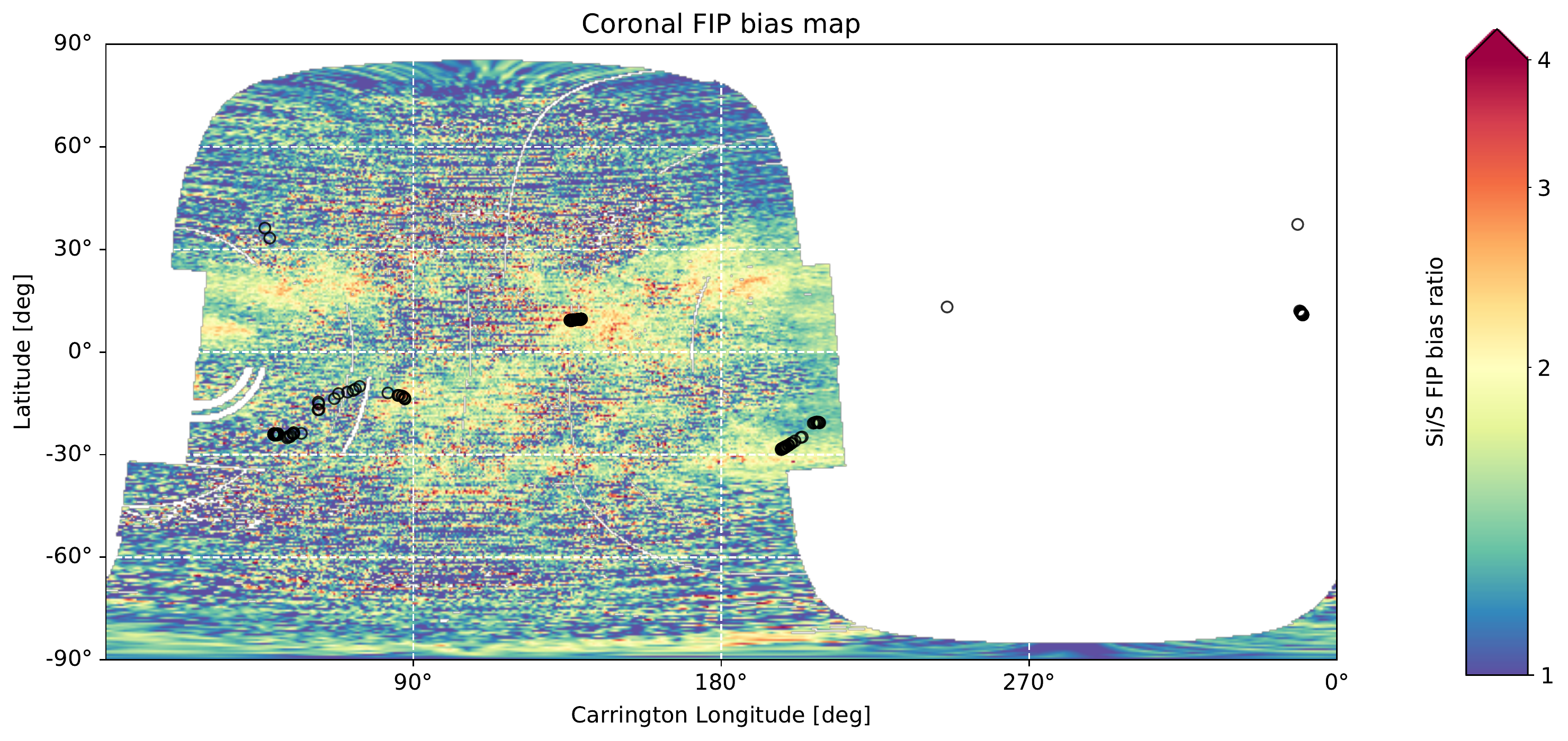}
	\includegraphics[width=2\columnwidth]{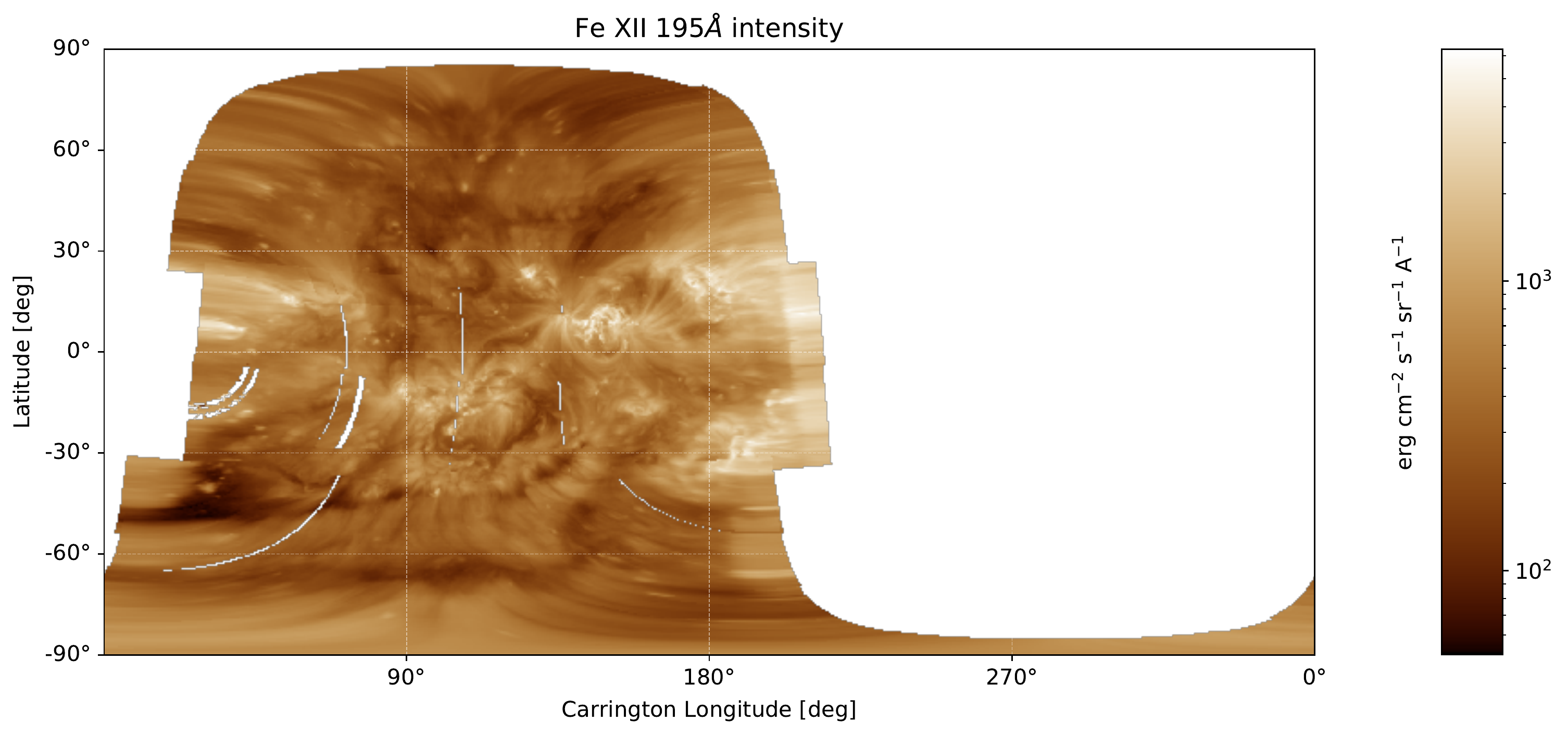}
	\caption{Synoptic maps showing FIP bias and EUV emission. The top panel shows a map of the Si/S First Ionisation Potential (FIP) bias ratio, with predicted ACE solar wind source regions overlain with black circles. The bottom panel shows total intensity in the FE XII 195.120\AA~line, which lies close to the AIA 193\AA~passband shown in Fig.~\ref{fig:ss_mapping}. The brighter areas of EUV emission, corresponding to active regions, can be seen to exhibit larger FIP biases.}
	\label{fig:fip_map}
\end{figure*}

In order to project the magnetic footpoints on to the composition maps derived from EIS data, each of the 26 individual EIS maps observed in a helioprojective frame (ie. solar X and solar Y, as seen by the telescope) was projected into a Heliographic Carrington frame (ie. longitude and latitude on the surface of the Sun). Each of the reprojected images were then added to form the final map. Where the images overlapped, the mean FIP bias in the overlapping pixels was taken. This FIP bias Carrington map is shown in the top panel of Fig.~\ref{fig:fip_map}, with the predicted solar wind source footpoints overplotted as circles. The bottom panel of Fig.~\ref{fig:fip_map} shows the EIS 195\AA~EUV intensity for context, where bad pixels in the original data show up as white lines. These bad pixels were straight vertical lines in the original helioprojective frame, but when transformed into the Carrington frame have been distorted into curves. Comparing the EUV intensity and FIP bias maps reveals a trend for higher FIP bias ratios in areas of brighter EUV emission, ie. active regions, as expected \citep{Brooks2015, Doschek2019}.

\begin{figure}
	\includegraphics[width=\columnwidth]{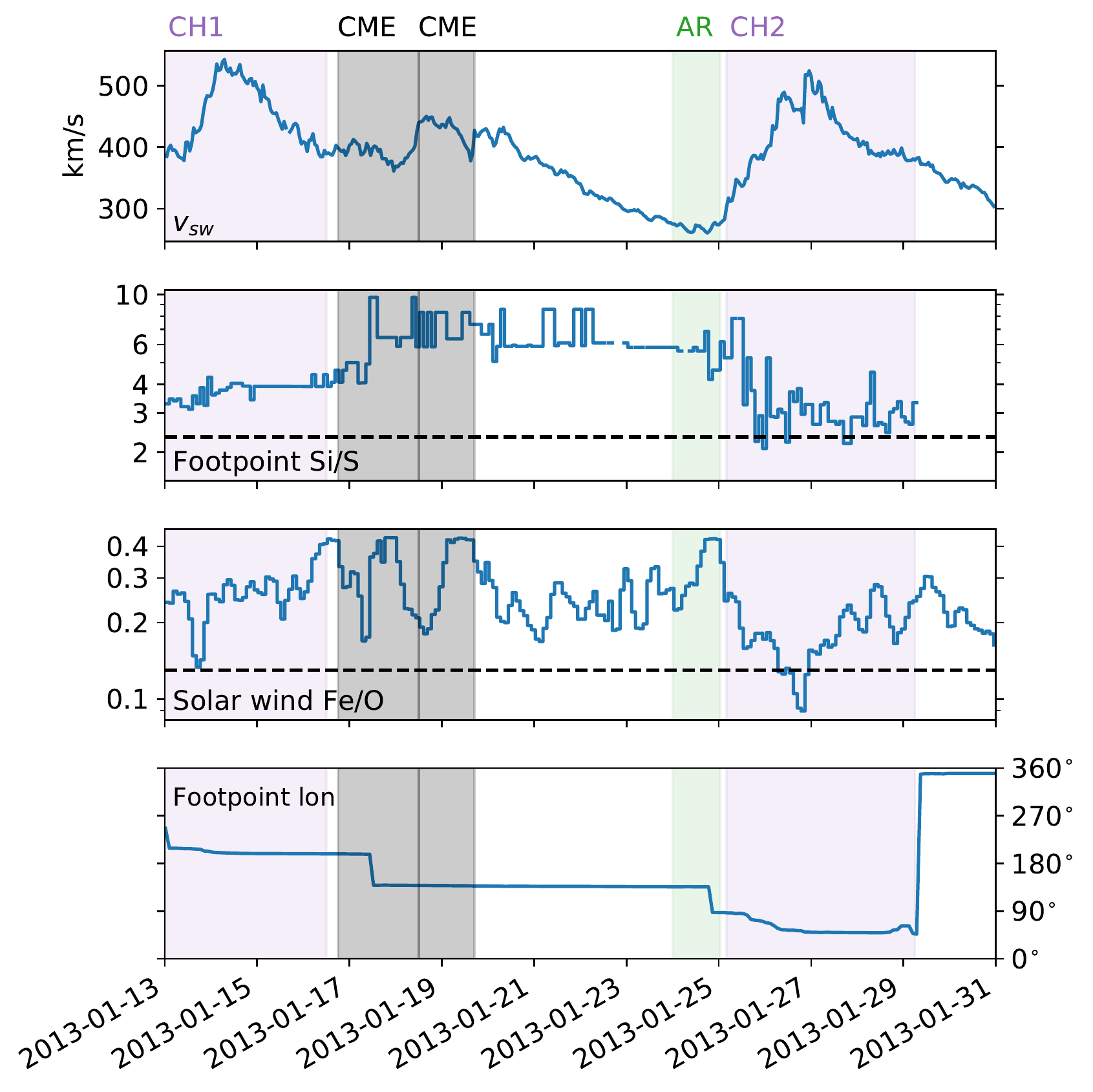}
	\caption{Direct comparison of inferred coronal Silicon to Sulphur coronal abundance (second panel) at the predicted solar wind source regions and solar wind Iron to Oxygen abundance (third panel). Dashed horizontal lines show photospheric abundances taken from \cite{Scott2015a, Scott2015}. The top panel shows in-situ measured solar wind speed for context, and the bottom panel shows the predicted footpoint Carrington longitude.}
	\label{fig:tseries_comparison}
\end{figure}

The coronal composition at the predicted sources of the solar wind measured by ACE can be extracted from the Carrington version of the full Sun composition map, by taking the FIP bias values at the predicted footpoints. Fig.~\ref{fig:tseries_comparison} shows a direct comparison between coronal footpoint composition and in-situ composition. The top panel again shows the solar wind speed for reference, with the second panel showing the coronal footpoint Si/S ratio. Across the whole interval the coronal Si/S values split into three regions, associated with three different solar wind sources. At the start of the interval (14 Jan 2013 to 18 Jan 2013) low Si/S ratios $\sim$~4 are associated with a coronal hole source. This then transitions to a predicted active region source (18 Jan 2013 to 25 Jan 2013) with high Si/S ratios $\sim$ 6, before transitioning back to a second coronal hole source (25 Jan 2013 to 30 Jan 2013) with lower Si/S values $\sim$2~--~3.

By contrast, the solar wind Fe/O ratio, shown in the third panel of Fig.~\ref{fig:tseries_comparison}, is highly variable during this time period. Despite this, (excluding the CME intervals) there are two clear dips to low Fe/O values $\sim$ 0.1, corresponding to relatively unfractionated plasma, during the coronal hole intervals. On daily timescales, the lack of a 1:1 correspondence between the in-situ measurements and remote measurements is not particularly surprising, due to inherent uncertainties in the mapping. This includes the limitations of the PFSS model, and the limitations of the ballistic backmapping.

\begin{figure}
	\includegraphics[width=\columnwidth]{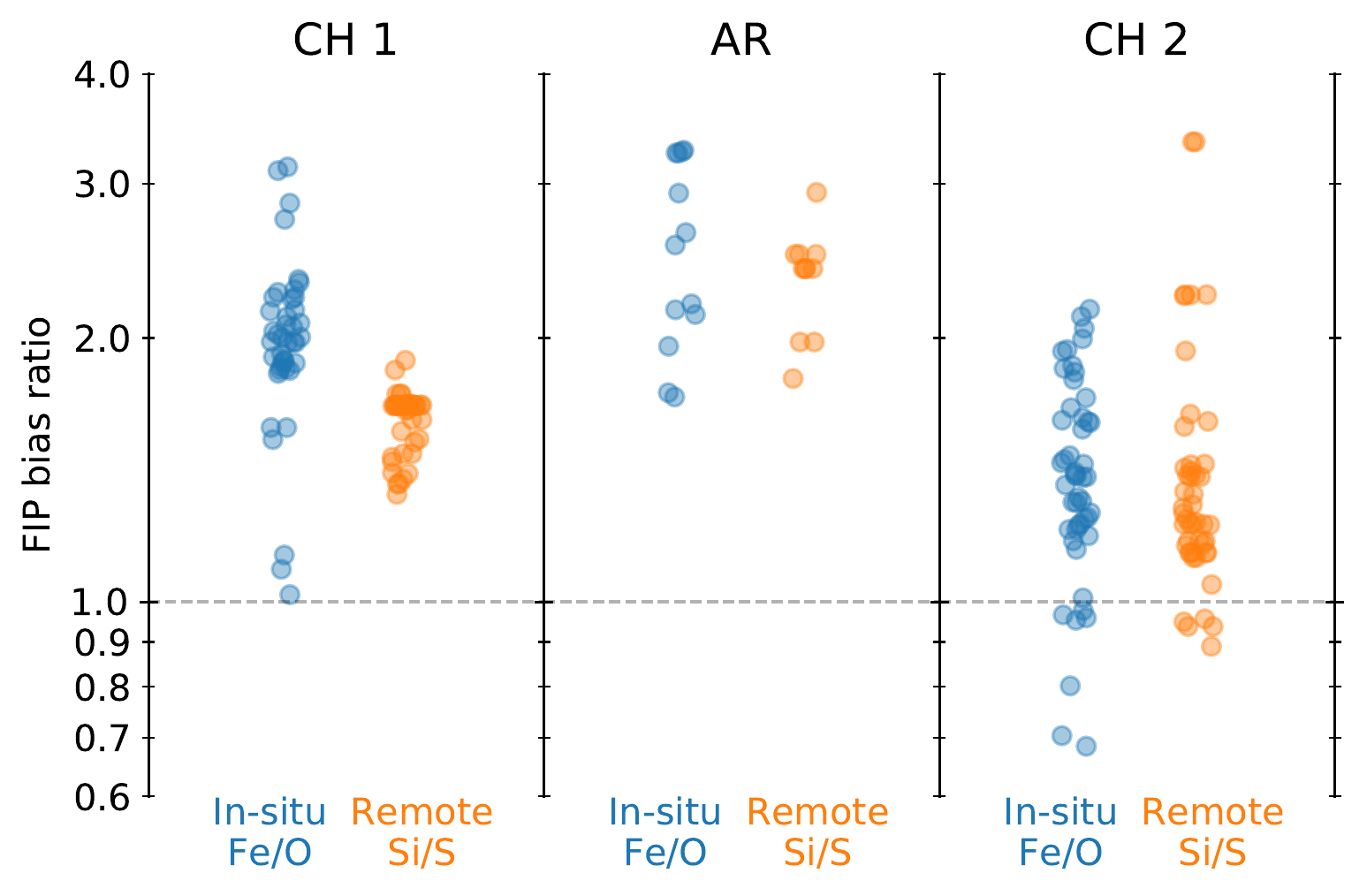}
	\caption{Distributions of FIP bias ratios in the three identified solar wind source regions. Solar wind Fe/O ratios are shown in blue, and coronal Si/S ratios are shown in orange.}
	\label{fig:boxplots}
\end{figure}

On larger time scales a difference should persist between different sources, as the mapping was earlier validated by comparing the model and in-situ magnetic field polarities. As such, Fig.~\ref{fig:boxplots} compares the distribution of solar wind and coronal compositions during the three intervals marked by bands in Figures \ref{fig:L1polarity} and \ref{fig:tseries_comparison}. On average, the mean FIP bias of each stream appears to behave in a similar way both remotely and in-situ, with the first coronal hole having on average intermediate values $\sim$2, the active region having larger values $>$2, and the second coronal hole having smaller values between 1 and 2.

There is a large spread in values in all three cases however, particularly within the two coronal holes. In pure coronal hole wind we would expect very little fractionation \citep{Feldman1998, Steiger2000}, and therefore FIP bias ratios close to 1. The in-situ observations show narrow time windows of around 6 hours where the coronal hole origin wind has ratios around 1, but for the majority of time the ratios are much higher. This is consistent with suggestions that fast wind emitted in a steady state in these coronal holes is relatively rare, and that interchange reconnection is responsible for the wind where the FIP bias is $>$~1 \citep[e.g.][]{Baker2009, Crooker2012b, Owens2020}. This is most probably because the prevelance of active regions during this interval at solar maximum disrupts the otherwise steady outflow of coronal hole wind \citep[e.g.][]{Macneil2019a}.

The distributions of FIP bias ratios for the three distinct sources overlap, meaning in the case studied here it is challenging to identify the type of solar wind source (coronal hole, active region) using only the in-situ composition measurements. This confirms recent statistical studies, that show the same overlap in distributions of in-situ elemental fractionation between different solar wind sources \citep{Zhao2017, Fu2017}. In our case this is particularly clear from the timeseries measurements of the two intervals of faster wind (Fig.~\ref{fig:tseries_comparison}), which show highly variable abundance ratios above the photospheric values.

\section{Discussion and conclusions}
\label{sec:conclusions}
We have presented a direct comparison of solar wind heavy ion composition and coronal composition at the predicted sources, over a period of 18 days at solar maximum. On daily timescales, within individual solar wind sources, the coronal and solar wind composition is poorly correlated (Fig.~\ref{fig:tseries_comparison}, middle panels), most likely because of the inherent uncertainties in the simple backmapping scheme used. On larger scales, the distribution of fractionation values within each distinct stream is similar and exhibits similar stream-to-stream variations in both the corona and solar wind (Fig.~\ref{fig:boxplots}). Although it is promising that the elemental fractionation from different sources matches, our method comes with some limitations. We are assuming that the coronal composition as measured by EIS did not change between the time of observation (16 -- 18th January) and the solar wind interval (13th -- 31st January). This is a limitation of the \emph{Hinode}/EIS observing plan, and in theory could be removed by a longer duration observation spanning two weeks. In addition the connection model (PFSS and ballistic back-mapping) could be replaced by more advanced heliospheric magneto-hydro-dynamic (MHD) models \citep[e.g.][]{Riley2011, Riley2019, Odstrcil2020}.

In the future, the methodology and data processing framework employed here can be used as a generic tool to perform more comparisons between coronal and solar wind plasma properties. For example, EIS also has the capability to measure coronal electron temperatures, densities, and mass fluxes \citep[e.g.][]{Brooks2015}, which could be compared to similar in-situ measurements in the near-sun solar wind \citep[e.g.][]{Halekas2020, Bercic2020, Macneil2020} to study the evolution of plasma as it transitions from the corona to the solar wind. 

Finally, our results have particular implications for the recently launched \emph{Solar Orbiter} misison \citep{Muller2012}, which carries both a coronal spectrometer \citep[SPICE,][]{TheSpiceConsortium2019} and solar wind plasma analysers (SWA, Owen et al., submitted to A\&A). These two instruments will allow for a replication of the methodology presented here (with SPICE replacing \emph{Hinode}/EIS, and SWA replacing ACE/SWICS). One of the key goals of \emph{Solar Orbiter} is exploiting links between remote and in-situ measurements to make new discoveries, and it has been envisioned that comparing remote and in-situ heavy ion composition will help facilitate this goal. The results shown here suggest that it will not be possible to use composition information to identify the spacecraft-corona connection on daily or smaller timescales. However, our demonstration that the distribution of fractionation values between different sources is consistent provides a new method to check a predicted connection.

\begin{acknowledgements}
We thank the anonymous reviewer and Alessandro Bemporad for helpful comments on the initial draft of the paper.

D.~Stansby, D.~Baker, and C.~J.~Owen are funded under STFC consolidated grant number ST/5000240/1. The work of D.~H.~Brooks was performed under contract to the Naval Research Laboratory and was funded by the NASA Hinode program.

This work utilises data from the National Solar Observatory Integrated Synoptic Program, which is operated by the Association of Universities for Research in Astronomy, under a cooperative agreement with the National Science Foundation and with additional financial support from the National Oceanic and Atmospheric Administration, the National Aeronautics and Space Administration, and the United States Air Force. The GONG network of instruments is hosted by the Big Bear Solar Observatory, High Altitude Observatory, Learmonth Solar Observatory, Udaipur Solar Observatory, Instituto de Astrof\'{\i}sica de Canarias, and Cerro Tololo Interamerican Observatory.

\emph{Hinode} is a Japanese mission developed and launched by ISAS/JAXA, collaborating with NAOJ as a domestic partner, NASA and UKSA as international partners. Scientific operation of the \emph{Hinode} mission is conducted by the \emph{Hinode} science team organised at ISAS/JAXA. This team mainly consists of scientists from institutes in the partner countries. Support for the post-launch operation is provided by JAXA and NAOJ (Japan), UKSA (U.K.), NASA, ESA, and NSC (Norway). 

The authors are grateful to the ACE, GONG, AIA, and Hinode/EIS instrument teams for producing and making the data used in this study publicly available. Data processing was carried out with the help of HelioPy v0.10.1 \citep{Stansby2020a}, astropy v4.0.1 \citep{TheAstropyCollaboration2018}, and SunPy v1.1.3 \citep{TheSunPyCommunity2020, Mumford2020}. Figures were produced using Matplotlib v3.1.3 \citep{Hunter2007, Caswell2020b}.

Code to reproduce the figures in this paper is available at \url{https://github.com/dstansby/publication-code}.
\end{acknowledgements}

\bibliographystyle{aa}
\bibliography{/Users/dstansby/Dropbox/zotero_library}

\begin{appendix}
\section{Correcting recent ACE SWICS data}
After 08/2011 the SWICS instrument on board ACE had to undergo an operational reconfiguration, and the number of heavy ions it could resolve was reduced. Both oxygen and iron charge states continued to be measured after this point, but the distribution of Fe/O abundance ratios in the new data is systematically higher than the pre-08/2011 dataset, and large values are clipped.
\begin{figure}
	\includegraphics[width=\columnwidth]{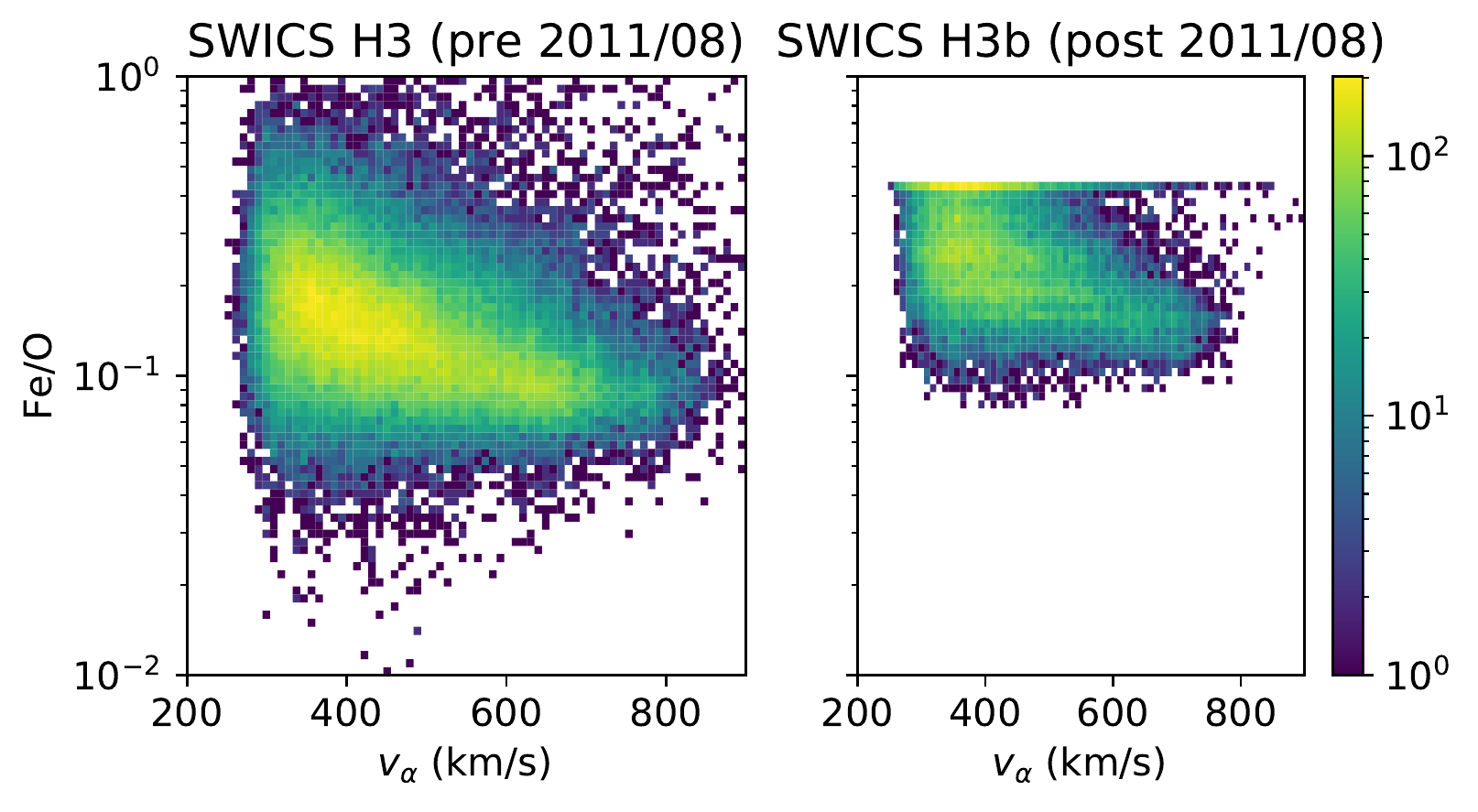}
	\caption{Comparison of the variation of Fe/O charge state ratio with velocity before (left-hand panel) and after (right-hand panel) the ACE SWICS reconfiguration in 08/2011.}
	\label{fig:swics}
\end{figure}
Figure \ref{fig:swics} shows the joint distribution of Fe/O ratio and alpha particle speed both before and after 08/2011. In the newer data, the ceiling is visible at Fe/O $\approx$ 0.4, and the distribution of charge states is systematically higher than the data before the re-configuration.

\begin{figure}
	\includegraphics[width=\columnwidth]{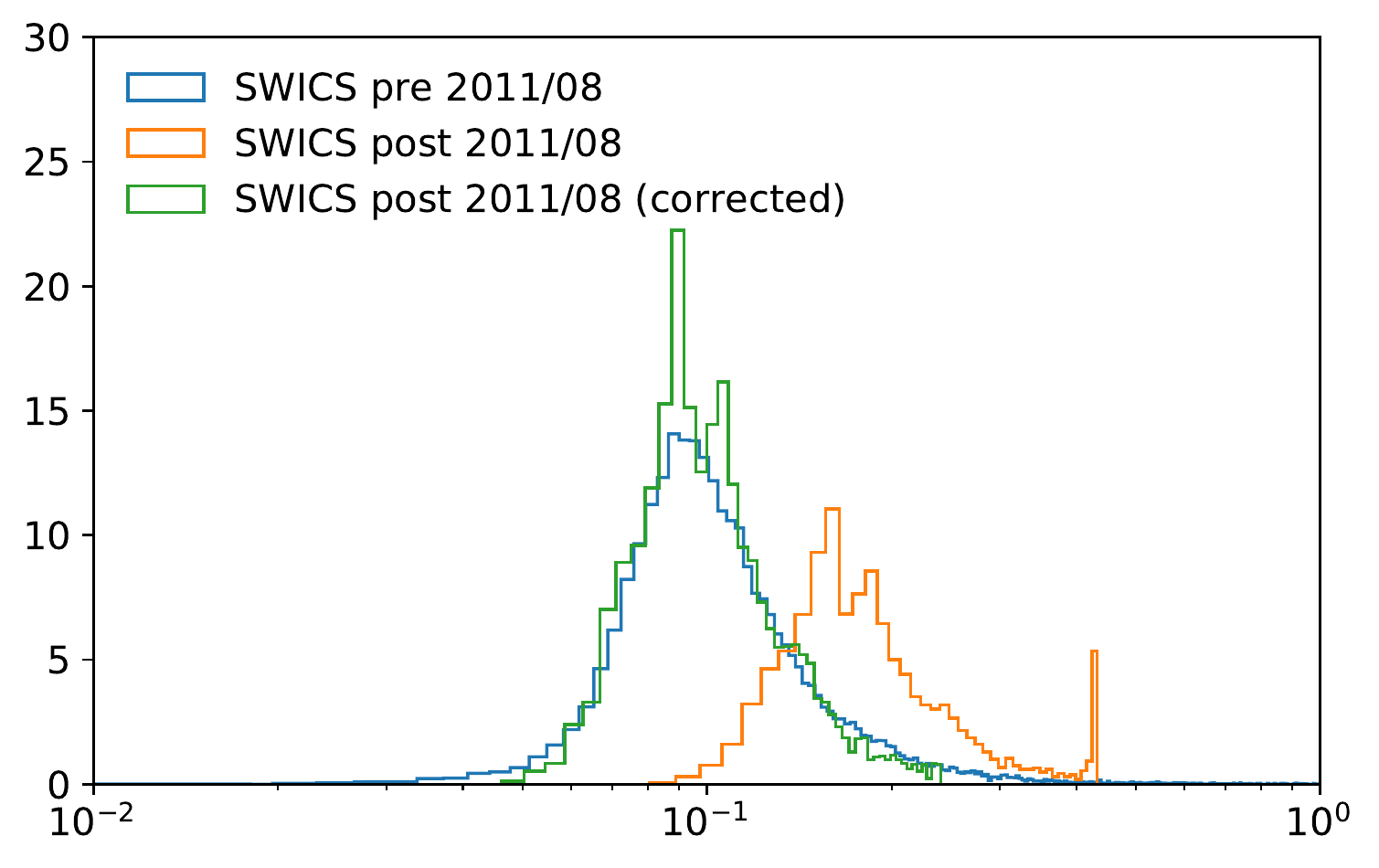}
	\caption{Histograms of Fe/O ratios in the range 600~km/s~$< v_{\alpha}<$~700~km/s before reconfiguration (blue) and after reconfiguration (orange). Corrected values for the post-reconfiguration data are shown in green.}
	\label{fig:swics_corrected}
\end{figure}

In order to correct for these differences, two steps were taken. First, data with Fe/O $>$ 0.42 in the new dataset were discarded. Secondly, the pre-configuration data were considered a ground truth, and a multiplicative constant between the old and new data estimated by matching the distributions of Fe/O values with alpha particle velocities in the 600 - 700 km/s speed range. The correction is
\begin{equation}
	\left ( \frac{Fe}{O} \right )_{corrected}  = 0.573 \left ( \frac{Fe}{O} \right )
\end{equation}
Figure \ref{fig:swics_corrected} shows distributions of Fe/O in this speed range, showing the original and post-reconfiguration data distribution, along with the empirically corrected values.

\end{appendix}
\end{document}